\pdfoutput=1
\documentclass[review]{elsarticle}
\usepackage{multirow}
\usepackage{threeparttable}
\usepackage{amssymb}
\usepackage[margin=2.9cm]{geometry}
\usepackage{hyperref}
\hypersetup{
    colorlinks=true,      
    urlcolor=blue,
    linkcolor=black,
    citecolor = black}

\begin{document}

\begin{frontmatter}

\title{About the consistency of the energy scales of past and present instruments detecting  cosmic rays above the ankle energy}

\author{Antonio Codino
}
\address{Department of Physics of Perugia University   and INFN, Italy}

\begin{abstract}
The energy spectrum of the cosmic radiation in the range 10$^{19}$-2.4$\times$10$^{21}$ eV has been recently predicted showing a rich and distinctive staircase profile. In order to check the prediction, the spectra measured by running and past experiments above 10$^{19}$ eV are examined. The computed spectrum compares more favourably with the Telescope Array, HiRes I and Yakutsk data rather than with the Auger data in the range (1-20)$\times$10$^{19}$ eV. Previous flux measurements by Haverah Park, {\sc sugar}, {\sc agasa} and Fly's Eye experiments are above the predicted spectrum in the limited band (1-30)$\times$10$^{19}$ eV. The flux measured by the Auger Group in the band (8-18)$\times$10$^{19}$ eV is below those of all other experiments and below the prediction.

  The energy scales of the instruments might be at the origin of the flux mismatch among the experiments. Accordingly, the energy scales of all the eleven instruments operating above 10$^{20}$ eV are examined and the major inconsistencies discerned. The paucity of events above 10$^{20}$ eV of the Auger experiment with respect to all others is by far the major puzzle emerging from this scrutiny. The Auger instrument recorded only 4 events above 10$^{20}$ eV with an exposure exceeding 42500 km$^{2}$ sr year while the Telescope Array recorded 13 events with an exposure of 8100 km$^{2}$ sr year. A tentative solution of this puzzle is ventilated.
\end{abstract}

\begin{keyword}
GZK effect\sep ultrahigh energy cosmic rays\sep cosmic ray spectrum
\end{keyword}

\end{frontmatter}

\section{Introduction}
An open problem of historical resonance in Cosmic Ray Physics is the comprehension of the energy spectra around 10$^{20}$ eV measured before the year 2003 by the of Haverah Park \cite{cunn80}, SUGAR \cite{winn86} and AGASA \cite{tak02,shino06} Collaborations compared to those measured after the year 2004 by HiRes \cite{abba04} Auger \cite{yama07} and Telescope Array (hereafter TA) \cite{mat11} Collaborations. According to the last three Groups the spectrum has an event suppression above (2.5-5)$\times$10$^{19}$ eV in face of a smooth extrapolation. In many papers of HiRes, Auger and TA Groups the spectral break has been quantified by an ultrasoft constant index $\gamma$ ranging from 3.5 up to 5.5, which clearly stands out from the spectral index of about 2.67 measured in the energy decade 3$\times$10$^{18}$-3$\times$10$^{19}$ eV. According to AGASA, SUGAR and Haverah Park experiments there is no break at all as explicitly stated in conclusive data analyses \cite{winn86,shino06,law91}. Table I gives the number of events above 10$^{20}$ eV observed by all experiments since 1960.

To-day (2017) the exact energy position of the break is elusive in spite of the claims on the precision and reliability of the energy scales of running instruments. The spectral break gauged by the deviation of the spectrum from a power law takes place around (2.6-3.0)$\times$10$^{19}$ eV in the early data samples of the Auger experiment \cite{schu09}, while in a recent paper of the TA Collaboration \cite{jui16} the position of the break is at 6.30x 10$^{19}$ eV. These numerical figures have to be compared with energy resolutions of the TA and Auger instruments in the range 15-25 per cent above 5.0x 10$^{18}$ eV. In this work the position of the spectral break is set at 2.6$\times$10$^{19}$ eV as discussed elsewhere \cite{cod13} and this particular energy is designated by E$_L$ where L on foot is for {\sc liga} (\textit{Lack of particle Injection to the Galactic Accelerator}) \cite{cod17a}. The maximum energy imparted to protons by the Galactic Accelerator is the physical meaning of the energy E$_L$. Above this energy the spectrum is thoroughly devoid of protons.

\begin{figure}
 \begin{center}
  \includegraphics[width=0.8\textwidth]{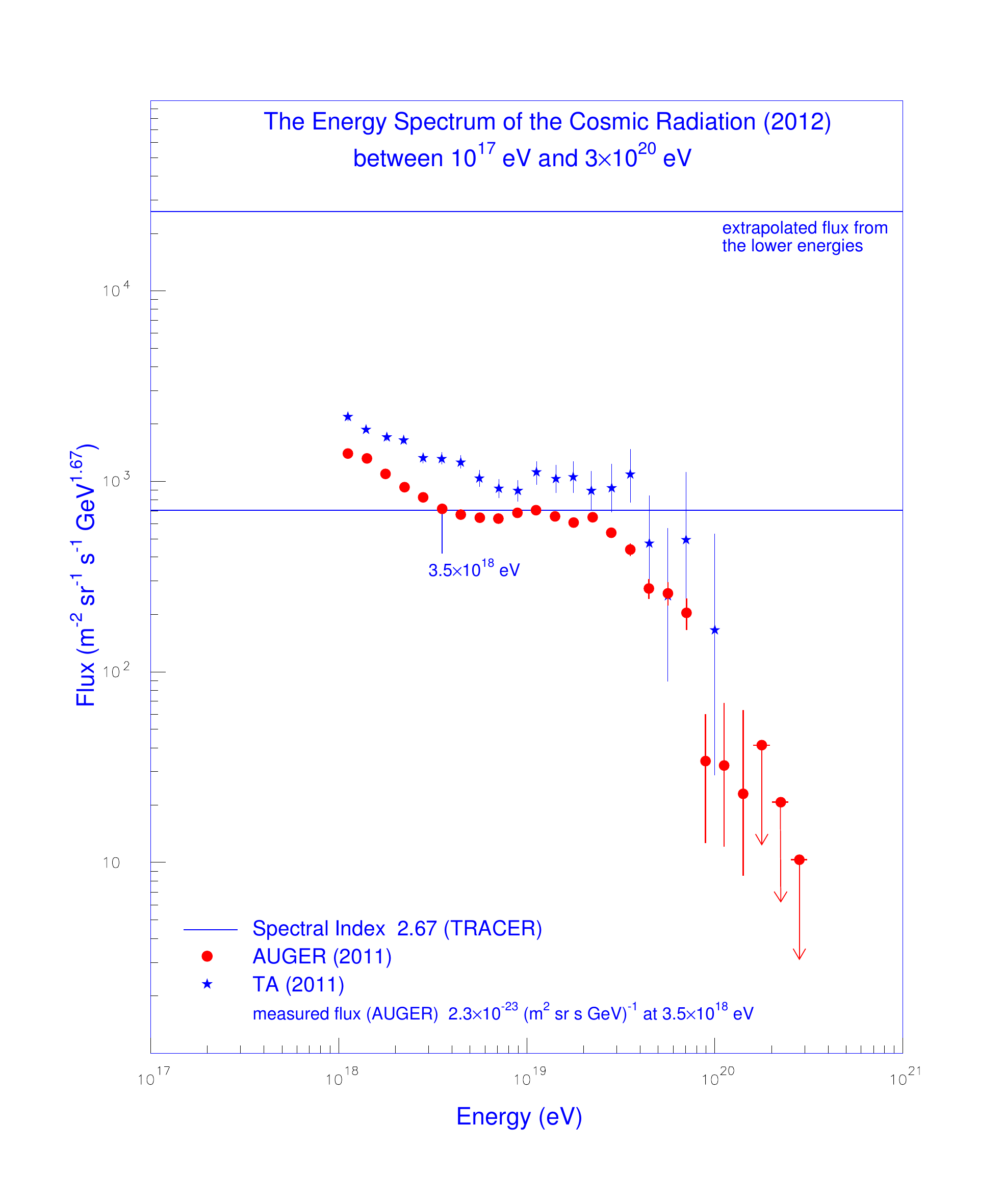}
\caption{Energy spectra of the cosmic radiation measured by the TA \cite{mat11} and Auger \cite{yama07} experiments.
There is a shift in the energy scales of about 15 per cent. \label{fig:fig1}}
\end{center}
 \end{figure}

An example of comparison of energy spectra measured by TA and Auger detectors is shown in fig. \ref{fig:fig1}. The two spectra exhibit a significant shift in intensity: the TA spectrum is more populated at the same energy E than the Auger spectrum. The flux mismatch persists in the last minute data of this year 2017.\\

\begin{figure}
  \begin{center} 
\includegraphics[width=0.8\textwidth]{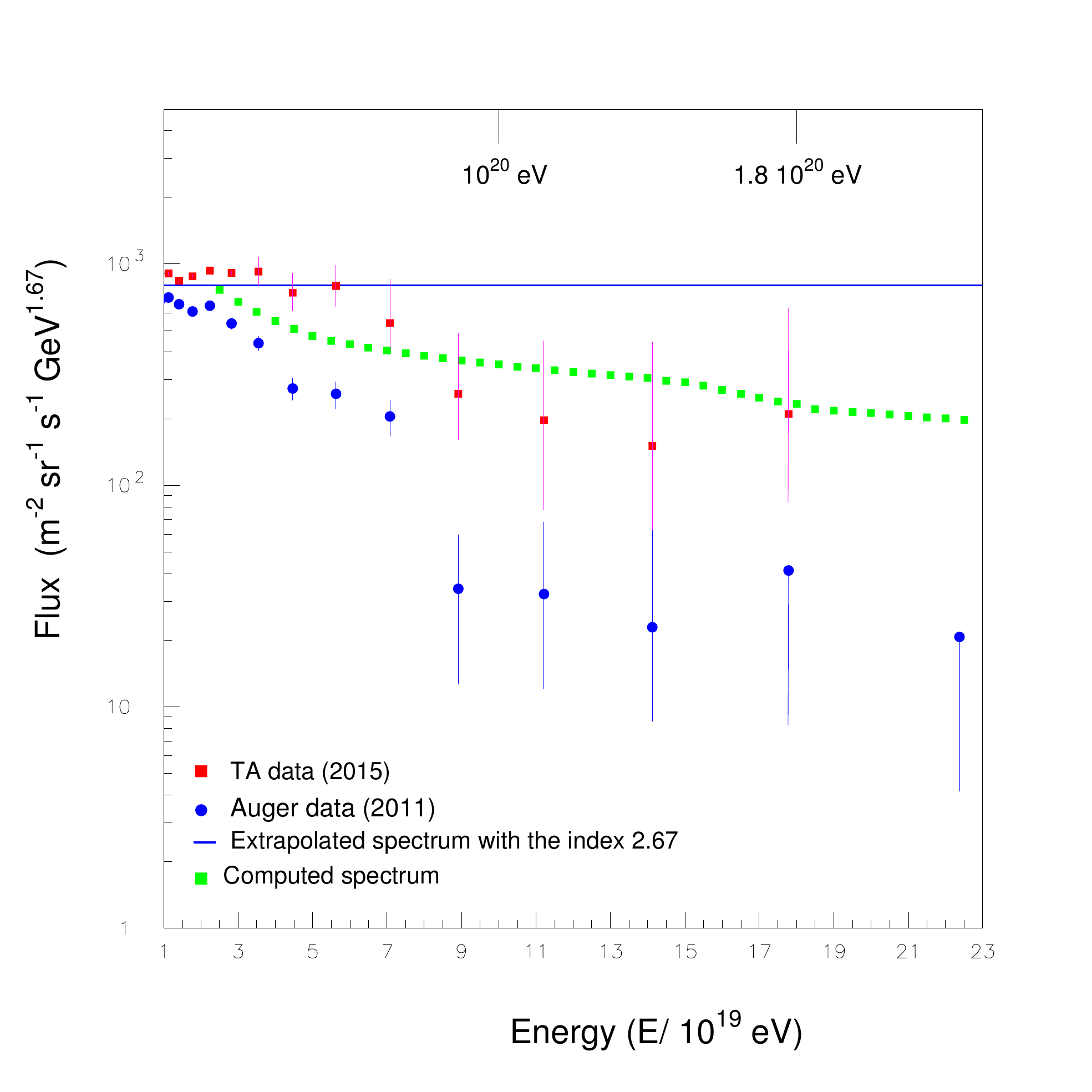}
\caption{Computed energy spectrum of the cosmic radiation (tiny green squares) in the interval 10$^{19}$-2.3$\times$10$^{20}$ eV expressed by the quantity JE$^{\gamma}$ where $\gamma$ is the spectral index and J the differential flux. Details of the calculation can be found elsewhere \cite{cod17a}. Experimental data are from Auger (blue dots) \cite{aab15} and TA (red dots) \cite{tin14} experiments. The horizontal blue line is an extrapolation of the spectrum JE$^{\gamma}$ = 798.17 particles/m$^2$ s sr GeV$^{1.67}$ with a constant $\gamma$ = 2.67 normalized to the Auger flux of 1.159$\times$10$^{-24}$ particles/m$^2$ s sr GeV at 10$^{19}$ eV \cite{aab15}. \label{fig:fig2}}
\end{center}
\end{figure}

The aim of these work on the extreme energy events of the cosmic radiation as they have been measured over many years by a number of detectors is to seek for a check and validation of a new calculation predicting the energy spectrum in the interval 10$^{19}$-2.4$\times$10$^{21}$ eV. The method and principles of the calculation has been recently described \cite{cod17a}, previously (2015) anticipated in a conference communication \cite{cod15a} and further substantiated \cite{cod17b}. The predicted spectrum in the limited range 10$^{19}$-2.3$\times$10$^{20}$ eV is reported in figure \ref{fig:fig2} (green tiny squares) and consists of very distinctive silhouette (flight of steps). This spectrum agrees with the Auger and TA data up to 9$\times$10$^{19}$ eV and with calorimetric measurements of Yakutsk (see fig. \ref{fig:fig6}) and Fly's Eye experiments. But patently, above 9$\times$10$^{19}$ eV the computed spectrum is incompatible with the Auger flux which is too small, being about 6.0$\times$10$^{-29}$ particles/m$^{2}$ s sr GeV above 10$^{20}$ eV. The flux deficiency of the Auger spectrum above (8-9) $\times$10$^{19}$ eV is the main motivation for this paper.
  \begin{table}
\caption{Experiments detecting cosmic rays above 10$^{19}$ eV \label{tab:prova_table}}
\scalebox{0.8}{%
\begin{threeparttable}
\begin{tabular}{c|lcccccc}
\hline
& Period & Collecting & Particles & Numbers of & Exposure & Average & Estimated$^{\ast}$\\
& of & Area & detected & events & km$^2$ sr year & Altitude & Flux at 10$^{19}$eV\\
& operation & km$^2$ & & above 10$^{20}$eV & & & $\times$10$^{-24}$ part./m$^2$ sr GeV\\
\hline
 Volcano\vspace{-0.08cm}&\multirow{2}{*}{1960 - 1965} & \multirow{2}{*}{20}&muons$^{\blacktriangle}$ and\vspace{-0.08cm}&\multirow{2}{*}{1}&\multirow{2}{*}{230}&\multirow{2}{*}{834 g/cm$^2$}&\multirow{2}{*}{$\sim$2.50}\\
Ranch& & & electrons & &&&\\
 Haverah\vspace{-0.06cm} & \multirow{2}{*}{1962 - 1987} & \multirow{2}{*}{12} & muons and\vspace{-0.06cm}& 8 (1991)\vspace{-0.06cm}& \multirow{2}{*}{400}& \multirow{2}{*}{220 m} & \multirow{2}{*}{2.22 (1981)}\\
Park&  & & electrons& 0 (2003)& & & \\
Sugar & 1968 - 1979 &87 & muons & 3 & 600 & 250 m& 1.90 (Sydney model)\\
\multirow{2}{*}{Agasa} & \multirow{2}{*}{1993 - 2005} & \multirow{2}{*}{102} &muons and\vspace{-0.08cm}&11 (2000)\vspace{-0.08cm}& \multirow{2}{*}{1620} & \multirow{2}{*}{900 m} & \multirow{2}{*}{3.10 (2002)}\\
&  & &electrons&7 (2003)&  &  & \\
 \multirow{3}{*}{Yakutsk} & \multirow{3}{*}{1974 - living} & \multirow{3}{*}{18} &Cerenkov photons\vspace{-0.04cm}&4 (2007)\vspace{-0.04cm}& 215 (1983)\vspace{-0.04cm}&\multirow{3}{*}{900 m}&2.5 (1985)\vspace{-0.04cm}\\
& & & muons and\vspace{-0.04cm}&0 (2017)\vspace{-0.04cm}&300 (2014)\vspace{-0.04cm}&&3.54 (2004)\vspace{-0.04cm}\\
& & & muons and\vspace{-0.04cm}&&&&26.4 (2017)\vspace{-0.04cm}\\
& & & electrons&&&&2.64 (2017)\\
\multirow{2}{*}{Fly's Eye} & \multirow{2}{*}{1982 - 1992} &  &Fluorescence\vspace{-0.1cm}&\multirow{2}{*}{4}&\multirow{2}{*}{$\sim$3200}&\multirow{2}{*}{850 m}&\multirow{2}{*}{2.23 (1994)}\\
&&  &photons&&&&\\
HiRes I\vspace{-0.07cm}&\multirow{2}{*}{1997 - 2006}& &Fluorescence\vspace{-0.07cm}&\multirow{2}{*}{3$^{\triangledown}$}&\multirow{2}{*}{$\sim$4500}&\multirow{2}{*}{850 m}&\multirow{2}{*}{1.80}\\
monocular&& &photons&&&&\\  
HiRes II\vspace{-0.05cm}& \multirow{2}{*}{1999 - 2006} & &Fluorescence\vspace{-0.05cm}&\multirow{2}{*}{0$^{\triangledown}$}&\multirow{2}{*}{$\sim$1500}&\multirow{2}{*}{850 m} &\multirow{2}{*}{2.05}\\
monocular&  & &photons&&& &\\
HiRes\vspace{-0.05cm}& \multirow{2}{*}{1999 - 2006} & &Fluorescence\vspace{-0.05cm}&\multirow{2}{*}{2}&\multirow{2}{*}{$\sim$2400}&\multirow{2}{*}{850 m}&\multirow{2}{*}{2.1}\\
Stereo& & &photons&&&&\\
 \multirow{3}{*}{Auger} & \multirow{3}{*}{2004 - living} &\multirow{3}{*}{3200}&Fluorescence\vspace{-0.06cm}&1 (2008)\vspace{-0.06cm}&12500 (2008)\vspace{-0.06cm}&&\multirow{3}{*}{1.60}\\
 & & &photons muons\vspace{-0.06cm}&4 (2015)\vspace{-0.06cm}&42000 (2015)\vspace{-0.06cm}&870 g/cm$^2$&\\
 & & &and electrons&4 (2017)&52500 (2017)&&\\
\multirow{3}{*}{TA}&\multirow{3}{*}{2008 - living}&\multirow{3}{*}{700}&Fluorescence\vspace{-0.08cm}&\multirow{3}{*}{13 (2017)}&  \multirow{3}{*}{8100 (2014)}&& \multirow{3}{*}{1.95}\\
 & & &photons muons\vspace{-0.08cm}& & &850 m& \\
 & & &and electrons& & & & \\
\hline
  \end{tabular}
\begin{tablenotes}[para,flushleft]
$^{\blacktriangle}$By muons and electrons is meant any charged particle of the shower including pions, kaons, protons and others.\\
$^{\ast}$Flux estimated by the Author of this paper based on the published spectra except in those cases where published tabulated fluxes exist.\\
$^{\triangledown}$Based on the spectrum reported in figure 4 of \cite{abba04} (Physical Review Letters).\\
\end{tablenotes}
\end{threeparttable}
}
  \end{table}

\section{Some features of the predicted energy spectrum above 10$^{19}$ \lowercase{e}V}
The normalization between predicted spectrum and observed spectra in fig. \ref{fig:fig2} uses Auger data \cite{aab15} at the energy of 1.122$\times$10$^{19}$ eV with a flux of 1.159$\times$10$^{-24}$ particles/m$^{2}$ s sr GeV. For comparison, the flux of the revised energy spectrum of the Yakutsk Array (hereafter YA) at 10$^{19}$ eV is 2.64$\times$10$^{-24}$ in the same unit \cite{sabo17}. This normalization does not entail any bias in the subsequent discussion. The extrapolated spectrum is multiplied by E$^{\gamma}$ with $\gamma$ = 2.67 and it is represented by the horizontal blue straight line in fig. \ref{fig:fig2} at the coordinate of 790.7 in the unit specified. This blue line is an extrapolation of the spectrum above E$_L$ and it is only a visual guide.

The energy spectrum is forged in the pre-acceleration phase by a particle filter excluding the lightest particles from the acceleration cycle according to the arguments developed in the preceding work \cite{cod17a}. The event suppression discovered by HiRes Collaboration \cite{abba04}, the fifth stigma of the cosmic-ray spectrum, is just the energy where the sieving process initiates to show up and refers to protons, the lightest nuclei. The lowest energy where the sieving process manifests itself is designated by E$_{LIGA}$ and concisely by E$_L$ as already noted. Once the value E$_L$ is correctly measured, the sieving points of all other cosmic nuclei are also assigned according to the rule E$_L$(Z) = Z x E$_L$ \cite{cod17a}. Numerically, with the position E$_L$ = 2.6$\times$10$^{19}$ eV \cite{cod13}, the break energy is 5.2$\times$10$^{19}$ eV for Helium, 1.56$\times$10$^{20}$ eV for Carbon and 2.4$\times$10$^{21}$ eV for Uranium. The corresponding intensities for H, He and C are, respectively, 1159.9, 112.62 and 3.795 in units of 10$^{-28}$ particles/m$^{2}$ s sr GeV and 744.8, 460.6, 291.6 in the units of figure \ref{fig:fig2}.

No nucleus of atomic number Z will compose the cosmic radiation above the energy Z x E$_L$. Accordingly, in the energy range above 5.2$\times$10$^{19}$ eV, quite amenable for TA and Auger instruments, protons and helium nuclei are thoroughly absent from the cosmic-ray spectrum.

Believing that the Auger energy scale in 2011 \cite{sala11} were more reliable than that of the incipient TA detector, the value E$_L$ = 2.6$\times$10$^{19}$ eV for protons has been adopted \cite[see fig. 4 ref.][]{cod15b}. The HiRes detector measured the break position at 5.5$\times$10$^{19}$ eV, which is very similar to that recently reported by TA Group \cite{fuku15}. In principle, since the maximum energy of a cosmic-ray event is 3$\times$10$^{20}$ eV as measured by the Fly's Eye Collaboration \cite{bird94,bird93}, the comparison of the predicted spectrum with the experimental data is feasible in the range 10$^{19}$- 3.0$\times$10$^{20}$ eV.

The silouhette of the energy spectrum (green squares, fig. \ref{fig:fig2}) and the associated chemical composition in the range 2.6$\times$10$^{19}$- 3.0$\times$10$^{20}$ eV are unique, and fortunately, thoroughly different from the features expected from the hypothetical GZK effect believed to materialize in the same energy range. Ultimately, the fictitious GZK effect is not observable in the Milky Way Galaxy for the simple reason that ultrahigh cosmic nuclei sprout, live and die in its interior. The plausibility of this statement, argued elsewhere \cite{cod15b,cod15c}, roots to facts, and not to toxic theories as those conceiving the GZK effect.

The energy spectra in this paper are expressed in a linear scale of energy, which is appropriate to the Galactic Accelerator \cite{cod17a,cod15b}, and the preferred unit of energy is 10$^{19}$ eV. 

\section{The energy spectra measured by the Auger and TA experiments}\label{sec:3auta}
The flux J at a given energy E is determined by J = N/ w AT where N is the number of observed events, w is the width of the energy bin, A is the instrument aperture and T the acquisition time. The exposure, e, is defined by e = A T and allows to compare the fluxes of instruments differing in collecting area and acquisition time. Generally the critical parameter in flux determination is the energy scale while aperture calculations seem to have less uncertainties and less biases. 
  Global traits of the energy spectra resulting from TA and Auger measurements are: 
\begin{itemize}
\item[(1)] The average flux measured by the Auger detector \cite{aab15} above and around 10$^{20}$ eV is about 
  6$\times$10$^{-29}$ particles/m$^{2}$ s sr GeV while that of the TA detector \cite{tin14} is about 38$\times$10$^{-29}$ in the 
  same unit. 
\item[(2)] In the energy decade 3$\times$10$^{18}$-3$\times$10$^{19}$ eV, featured by a constant index $\gamma$ = 2.67 \cite{aab15}, the average fluxes of TA and Auger experiments are, respectively, about 1000 and 800 particles/m$^{2}$ s sr GeV$^{1.67}$. Accordingly, the flux gap is: ( 1000. - 800.)/ 1000. = 20 per cent. This difference can be considered as constant and it is a solid, reliable reference to scrutinize the energy scales in the region where the flux gap enlarges beyond the systematic uncertainties, e.g. the interval (5-20)$\times$10$^{19}$ eV. 
\item[(3)] The flux resulting from inclined cascades measured by the Auger detector in 2015 \cite{aab15} 
  does not clarify the flux gap problem of fig. \ref{fig:fig1} in the region (1-30)$\times$10$^{19}$ eV. The inclined
  spectrum has a precious, intrinsic value but it is irrelevant for solving the flux gap problem.
  \textit{De facto} the inclined spectrum observed by Auger instrument enhances the confusion on the
  true energy scale of the most energetic events. This is evident from the error bars of the
  inclined spectrum which diverge above 5$\times$10$^{19}$ eV (see fig. 5 of ref. \cite{aab15}). 
\item[(4)] No obvious instrumental reason could explain the flux gap problem shown in figure \ref{fig:fig1} because
   any plausible parameters affecting the systematic and statistical errors have already been
   reconnoitred by the TA and Auger Groups and by consortia \cite{maris14,daw13}. 
\end{itemize}
A priori, why the TA and Auger energy scales have to coincide within the systematic errors and not to diverge further? The consortium of fractions of TA and Auger Groups \cite{maris14} aiming at to alleviate the flux gap problem in 2014 proposed a rigid shift of the energy scales by $\pm$ 8 per cent. With this operation the TA and Auger energy spectra in the very limited region (0.1-5)$\times$10$^{19}$ might overlap but the major portion of the spectrum (5-20)$\times$10$^{19}$ eV would still suffer a large mismatch. Any rigid shift of the energy scales does not resolve the flux gap shown in fig. \ref{fig:fig1}. Further support to this assertion is discussed in Section \ref{sec:6cal}.

Dominant sources of the systematic error in the energy scales of Auger and TA detectors are: 
(a) florescence yield of charged particles of the cascade ionizing air nitrogen; (b) Calibration of the telescope by artificial ultraviolet sources (photometric calibration). (c) air pollution level which affects the florescence light transmission down to the telescope. Clouds, dust grains, smokes, ozone and chemicals are major pollutants in the band 350-470 nm. (d) Cerenkov light released by the air cascades which has to be subtracted from the florescence light output; (e) invisible energy of the cascade determined by calculations of atmospheric cascade features; this energy amounts to about 8-10 per cent around 10$^{20}$ eV according to updated cascade simulations \cite{maria11}.

The systematic uncertainty in the energy scale of the Auger detector was 22 per cent \cite{yama07} in the band 10$^{18}$-10$^{20}$ eV and presently is 14 per cent while that of the TA detector about 21 per cent in the interval 10$^{19}$-10$^{20}$ eV. A global update of the energy scale of the Auger instrument was discussed in 2013 \cite{verzi}. For comparison the energy resolutions of AGASA \cite{tak03}, HiRes \cite{abba05} and YA \cite{glush14} experiments are, respectively, 18, 17 and 25 per cent. These numerical figures have been evaluated by the quoted experimental Groups.\\

In the case of the Auger and TA telescopes the dominant error sources (a) and (b) explain, respectively, 61 and 50 per cent of the systematic errors.

\section{The energy spectra measured by Haverah Park, Fly's Eye and AGASA experiments}
In the preceding section the fluxes of two similar, modern, unsurpassed detectors have been compared resulting in an abundant flux mismatch above 8$\times$10$^{19}$ eV, which renders the comparison between data and prediction of fig. \ref{fig:fig2} not stringent. Here the examination focuses on the fluxes measured by archaic instruments operated by the Haverah Park (1962-1987) and AGASA (1993-2005) Groups which reconstructs shower features only by ground charged particles and not by atmospheric florescence light. In spite of the term used above, archaic detectors here do not allude to incorrect or superseded measurements.

It turns out that: 
\begin{itemize}
\item[(5)] The fluxes measured by Haverah Park and AGASA experiments above 10$^{20}$ eV outnumber that measured by the Auger detector \cite{aab15} by more than one and two orders of magnitude being, respectively, 1370$\times$10$^{-29}$ \cite{law91} and 192 $\times$10$^{-29}$ \cite{tak02} particles/m$^{2}$ s sr GeV. 
\end{itemize}
  Above 10$^{20}$ eV the Volcano Ranch \cite{lin61}, Haverah Park \cite{law91} and SUGAR \cite{winn86} experiments recorded, respectively, 1, 8 and 3 events with exposures of 221, 600 and 245 km$^{2}$ sr year. The 8 Haverah Park events were subdivided into 4 vertical and 4 inclined events.\\ 

 Both AGASA and Haverah Park detectors sampled secondary charged particles of atmospheric cascades at ground. It is believed that the density of such particles at about 600 meters from the cascade core is proportional to the primary nucleus energy and almost independent on the nuclear species up to 3$\times$10$^{20}$ eV. This credence seemed to rely upon empirical evidence \cite{bow83} but the independence from the nuclear species is surprising and strictly incorrect as follows from the analytical theories of atmospheric cascades ( N$_{\mu}$ = k A$^{\beta}$ E$^{1- \beta}$ where k is a constant, A is the mass number of the primary, $\beta$ is the inelasticity in the range 0.13-1.15, E the energy of the primary nucleus and N$_{\mu}$ the number of muons in the cascade above a specified threshold ).\\
 
 The Haverah Park detector acquired data in the period 1968-1987 and the Group made the data analysis \cite{cunn80}, a conclusive analysis \cite{law91}, and hopefully, a terminating analysis \cite{ave02} with an energy estimator tuned with the CORSIKA package believed to be more adequate than the previously adopted energy estimators. The tangible result of the terminating analysis of the Haverah Park experiment is that the 4 vertical events above 10$^{20}$ eV, all descended below 10$^{20}$ eV. This miracle is just a byproduct of the modern simulation codes of air cascades and happened 15 years after the closure of the Haverah Park facility, simultaneously to the discovery \cite{abba04} of the spectral break at 2.6$\times$10$^{19}$ eV. Skepticism on the use of the radial function of charged particles of air cascades adopted by the Haverah Park Group to determine primary energies was vented by others \cite{khris85}.

\begin{itemize}
\item[(6)] The most energetic cosmic-ray event recorded by Fly's Eye detector \cite{bird94,bird93} in 1991 has an
  energy of (3.0 +0.36 - 0.54) $\times$10$^{20}$ eV. This event has never been disclaimed and implies a
   very high cosmic-ray flux above 10$^{20}$ eV. Its coordinates in fig. \ref{fig:fig2} are: 30 (energy) and 1365
  (flux). The average flux estimated by Fly's Eye is 31.0$\times$10$^{-29}$ particles/m$^{2}$ s sr GeV \cite{bird94}.
\end{itemize}

  Notice again that the flux of about 6.0$\times$10$^{-29}$ particles/m$^{2}$ s sr GeV observed by the Auger experiment above 10$^{20}$ eV (see fig. \ref{fig:fig2}) is well below that implied by the most energetic event recorded by Fly's Eye detector.\\ 

  At this step a trivial but necessary remark is that if the Auger flux around 10$^{20}$ eV is basically correct, the fluxes of all other experiments are badly measured and incorrect, including those of the TA and YA Groups. However, the reverse might be true, in spite of the superior technical capability of the Auger instrument; it detects both florescence light and charged particles on the ground in a huge arena, a fraction of 0.001156 of the continental Argentina.

  It is anticipated here that the Auger energy spectrum, besides the discrepancy with all other experiments, has intrinsic inconsistencies which emerge, for example, by charting the number of events above 10$^{20}$ eV versus exposure from 2004 to 2017 as scrutinized in the next sections.\\ 
 
  A restful note is that the flux gap problem in fig. \ref{fig:fig1} mitigates, by merely remembering that wide discrepancies in flux measurements using atmospheric showers are not new: for example CASA BLANCA and DICE experiments did measure fluxes lower than a factor of 2 or more in face of other experiments in energy bands below 10$^{16}$ eV. 
\section{What is at stake with the energy scales of Auger and TA experiments} \label{sec:5stake}
The equalization of the two cosmic ray fluxes of the vast TA and Auger instruments within their systematic errors is not only a superb technical challenge but it directly touches fundamental phenomena regarding Cosmology, Radio Astronomy and Cosmic Ray Physics.\\

The second fundamental piece of empirical evidence contradicting the interpretation of the spectral break discovered by the HiRes Group \cite{abba04} in terms of GZK effect derives from the energy scale \cite{cod13}. In fact, the deviation from a power-law extrapolation of the cosmic-ray spectrum reported in 2007 by the Auger Collaboration \cite{yama07} takes place between 2.5$\times$10$^{19}$ eV and 3.0$\times$10$^{19}$ eV, not above this energy band. According to this study and preceding ones \cite{cod13,cod17a} on the GZK theme, the incipient deviation from a power-law extrapolation of the spectrum is localized at E$_L$ = 2.6$\times$10$^{19}$ eV \cite[see fig. \ref{fig:fig4} of ref.][]{cod15b} utilizing the Auger energy spectrum of the year 2011 \cite{sala11}. Since the systematic error in the Auger energy scale amounts to 20 percent around 10$^{19}$ eV, the value E$_L$ = 2.6$\times$10$^{19}$ eV could ascend up to 3.2$\times$10$^{19}$ eV, not beyond this limit. The expected energy region for the GZK effect is above 6.0$\times$10$^{19}$ and the maximum hypothetical depression would lie around 2.1$\times$10$^{20}$ eV \cite{cod13}. In these numerical figures are the terms of the second inconsistency in the interpretation of the event suppression \cite{abba04} via the fictitious GZK effect as discussed in a previous work \cite{cod13}.

However, along the years the break energies determined by the Auger Group have become E$_L$ = (2.9$\pm$ 0.20)$\times$10$^{19}$ \cite{schu09}, E$_L$ = (4.26 $\pm$ 0.20)$\times$10$^{19}$ \cite{sala11} and E$_L$ = (5.24 $\pm$ 0.38)$\times$10$^{19}$ \cite{shu13}.\\

In short, during the period 2007-2016, the energy scale of the Auger detector drifted toward ascending values. Notice that the systematic uncertainty in the energy scale does not affect the {\sc liga} energies quoted above because they are results of the same instrument.\\

By shifting upward the energy scale of the early (2007) Auger instrument, e. g. repositioning the same events at adequate higher energies, of course implies the displacement of the break energy E$_L$ toward a critical limit where the {\sc liga} effect is disguised as GZK effect.

But what is definitely embarrassing to everybody is that the upward shift of the energy scale, by more than the nominal uncertainty of about 20 percent, has become to a downward shift around the energy of 2$\times$10$^{20}$ eV ! In fact the most energetic event detected by the Auger Observatory in the period 2004-2007 had an energy of 1.77$\times$10$^{20}$ eV \cite{yama07} but this energy was rescaled downward at 1.4$\times$10$^{20}$ eV \cite{sala11,aab15} without any reason whatever. In this way the energy scale does not suffer a rigid shift but a compressing stage, similarly while playing an accordion in the compressing phase, because the two extreme energies contract!

By maiming the extreme energy events above 10$^{20}$ eV the interpolation of the cosmic-ray spectrum via an ultrasoft index such as 4.5 or 5.5 becomes viable avoiding an immediate rejection of the GZK effect by ritual statistical tests.
\begin{figure}
 \begin{center}
\includegraphics[width=0.8\textwidth]{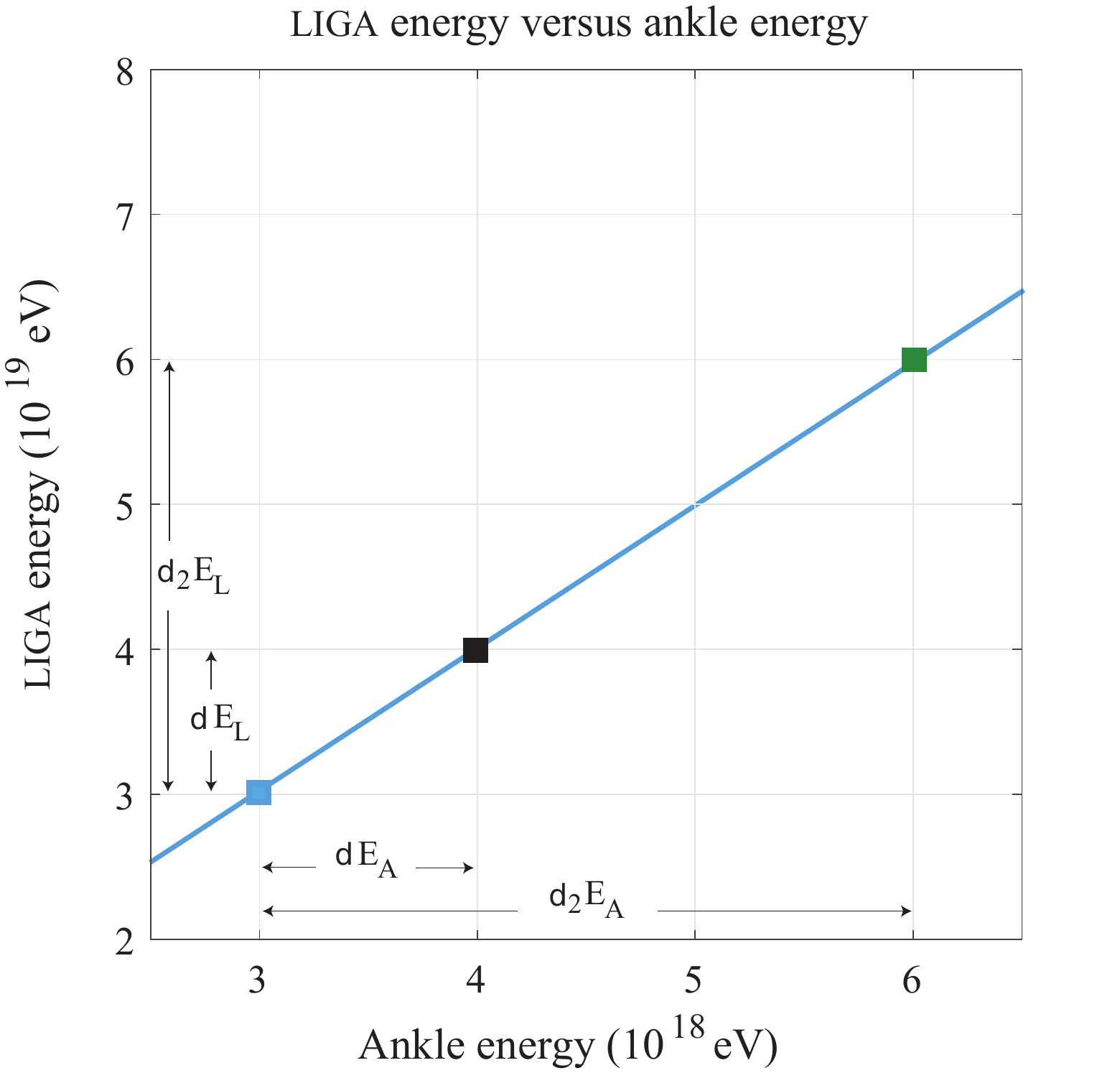}
\caption{Energy calibration of an ideal instrument using the ankle and {\sc liga} energies. The turquoise square represents one measurement of E$_L$ and E$_A$ energies, respectively, at the arbitrary, fictitious positions 3$\times$10$^{18}$ eV and 3$\times$10$^{19}$ eV. If the instrument for any reason undergoes an energy shift $\delta$E$_L$ in the vertical axis ({\sc liga} energy), the corresponding ankle energy will be displaced by $\delta$E$_A$  so that the new coordinates of the measurement are: E$_A$ + $\delta$E$_A$, E$_L$ + $\delta$E$_L$ (black square). Obviously, a similar behaviour is expected for a larger shift $\delta_{2}$E$_L$ (green square). \label{fig:fig3}}
\end{center}
 \end{figure} 
\section{Calibration of the energy scales by the {\sc liga} and ankle energies} \label{sec:6cal}
The examination of the energy scales of Fly's Eye, HiRes monocular, HiRes stereo and TA detectors in the energy interval 5$\times$10$^{17}$-3$\times$10$^{20}$ eV requires an adequate tool. This tool is a plane (see fig. \ref{fig:fig3}) where the vertical axis is the {\sc liga} energy, E$_L$, and the horizontal axis is the ankle energy, E$_A$.

Suppose that the event suppression \cite{abba04} characterized by the energy E$_L$ does really exist ignoring the intriguing results of the data analyses of archaic detectors (AGASA, Haverah Park and SUGAR Groups). In this way \textit{a priori} the doubts on the existence of the break are removed and only the exact value of the energy E$_L$ has to be measured and comfortably assessed. The measurements need precise and calibrated energy scales, or equivalently, small statistical and systematic errors.

The fourth stigma of the energy spectrum is the ankle which is located at the energy E$_A$ = 3.1$\times$10$^{18}$ eV according to precision measurements of Fly's Eye experiment \cite{bird94,bird93}. The cosmic-ray spectrum comprised between E$_A$ and E$_L$ is amenable for energy calibration because is featured by a constant spectral index $\gamma$ = 2.67 established by all experiments. The difference between the ankle energy and the {\sc liga} energy are milestones, real basements, bi-pillars for energy calibration. 

Suppose that the E$_L$ energy measured by an instrument via atmospheric cascades, due to the imperfect algorithms for energy assignment, results in a too high value (for example, 6$\times$10$^{19}$ eV) with respect to the true value (imagine, for example, E$_L$ = 2.6$\times$10$^{19}$ eV). In this condition the energy scale of the instrument is erroneously shifted upward, and consequently, the ankle energy E$_A$ will be dragged on toward unreal high values as well (rigid shift hypothesis of the energy scale).
 Fig. \ref{fig:fig3} shows an E$_L$ E$_A$ plane with a turquoise line arbitrarily positioned at an angle of 45 degrees. The turquoise line serves to comprehend the following reasoning.\\ 

\begin{figure}
\begin{center}  
\includegraphics[width=0.8\textwidth]{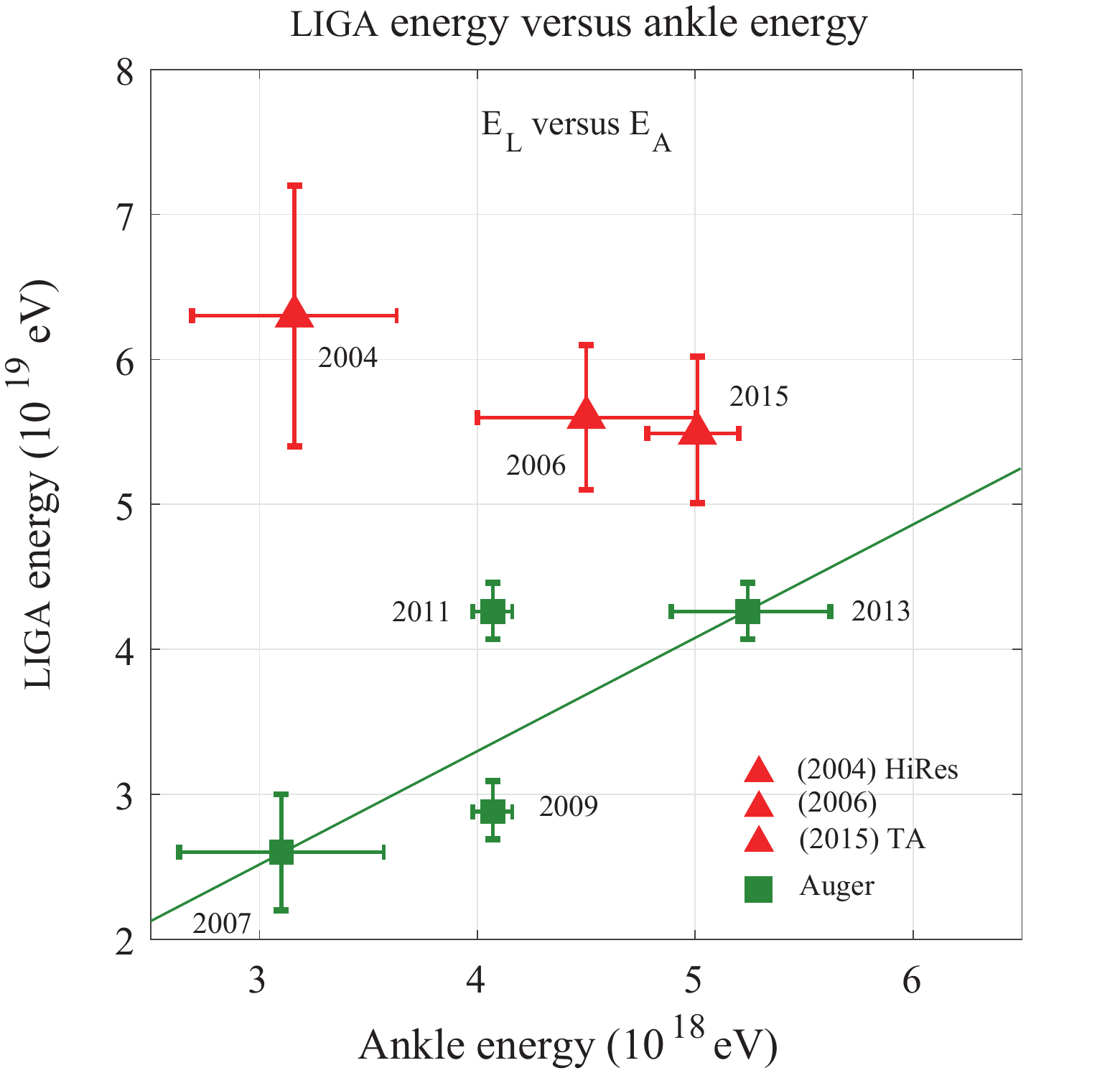}
\caption{Measurements of the {\sc liga} and ankle energies by Fly's Eye, HiRes, Telescope Array and Auger experiments. Auger data have a systematic error of about 22 per cent \cite{yama07}. The relative importance of statistical and systematic errors of the Auger data flashes in the comparison between the 2009 datum (only the statistical error is shown) and 2013 data point (only systematic error). The Auger data point labelled 2011 exhibits only a statistical error as well. The {\sc liga} energy of the data point labelled 2007 with E$_L$ = 2.6$\times$10$^{19}$ eV is based on the simple interpretation of the Auger energy spectrum \cite{yama07} made elsewhere \cite{cod13}. References of the data are given in the text. \label{fig:fig4}}
\end{center}
\end{figure}

  The calibration marks E$_A$ and E$_L$ do exist in the cosmic-ray spectrum\footnote{Presently the existence of the ankle is not questioned in Cosmic Ray Physics. The discovery of the event suppression in 2004 \cite{abba04} emerged in an extremely harsh scientific environment forged by the both devious and legitimate outcomes of AGASA \cite{tak02,shino06}, Haverah Park \cite{cunn80} and SUGAR \cite{winn86} experiments, which denied the fictitious GZK effect with inflated energy scales above 10$^{20}$ eV. It is conceivable that, in this environment, the major instrumental effort of the High Resolution Fly's Eye Collaboration was directed to the discovery of the spectral break, with the energy scale at hand, and not with an undisputable one. In the subsequent years, with the horizon cleared of the polemical clouds on the existence of the spectral break, the young (2004-2009) Auger instrument gauged the break energy at the coordinates 3.1$\times$10$^{18}$ eV and 2.6$\times$10$^{19}$ eV in the plane of fig. \ref{fig:fig3} ( bottom green square). According to the data of Fig. \ref{fig:fig5} and \ref{fig:fig6} the Auger Group after the year 2011 performed moot measurements of both E$_A$ and E$_L$ with adulterated energy scales as described in Sections \ref{sec:5stake} and \ref{sec:7drift}. 
    A forthcoming work will describe in detail how the interpretation of the spectral break \cite{abba04} in terms of {\sc liga} effect entails radical changes not only in Cosmology, Radio Astronomy and Cosmic Ray Physics but in all macroscopic sciences including Solar Physics.}. The difference (E$_L$ - E$_A$ ) is a measurable physical quantity. Measurements of E$_L$ and E$_A$ by a sole instrument will lie on the turquoise line, and if the energy measurements are simply correct, they will occupy a single, unique, unmovable point on the turquoise line. For clarity this imaginary point is materialized by a turquoise square in fig. \ref{fig:fig3} at the arbitrary coordinates E$_A$ = 3.0$\times$10$^{18}$ eV and  E$_L$ = 3$\times$10$^{19}$ eV. If the instrument, for any reason, suffers a tiny shift $\delta$ E$_L$ in the energy scale, the new {\sc liga} energy is E$_L$ + $\delta$ E$_L$ and the new ankle energy will be dragged on to the value E$_A$ + $\delta$ E$_A$ as well (black square). An additional shift $\delta_{2}$E$_L$ would displace the initial point ( turquoise square) to the point E$_L$ + $\delta_{2}$E$_L$ of figure \ref{fig:fig3} (green square). In this example all measurements of the {\sc liga} and ankle energies will lie on the turquoise line (fig. \ref{fig:fig4}) because of the rigid shift hypothesis. Concisely and in practical terms, such an imaginary instrument performs correct measurements but has an offset.

  Figure \ref{fig:fig4} reports the {\sc liga} energy versus ankle energy, that is E$_L$ versus E$_A$, measured by Fly's Eye, HiRes, TA (red triangles) and Auger experiments (green squares). The initial precise value of E$_A$ = 3.1$\times$10$^{18}$ eV measured in the year 1994 by the Fly's Eye Group \cite{bird94,bird93}, became (5.0 $\pm$ 0.2)$\times$10$^{18}$ eV in 2015 according to the TA Group \cite{fuku15}. The ascension of the ankle energy: (3.1 - 5.0 ) /3.1 = 61 per cent, outnumber the systematic error of about 20 per cent of Fly's Eye and TA experiments\footnote{The very core of the Telescope Array Collaboration are descendants of the HiRes experiment who are complemented with fragments of the AGASA Group, and of course, with new members. Presumably, this unique blend of minds is an excellent unbiased resource to perform correct measurements around 10$^{20}$ eV since they claimed \cite{abba04} and disclaimed \cite{tak02} at will the existence of the spectral break \cite{abba04}. The description of the TA instrument is disseminated in many papers (see for example \cite{barci11}).}.\\
  
  The green line in fig. \ref{fig:fig4} connects the first \cite{yama07} and the last \cite{shu13} data points of the Auger instrument measuring E$_A$ and E$_L$. The terms first and last refer to the times of measurement, or equivalently, to the minimum and maximum values of the ankle energies. The green line in fig. \ref{fig:fig4} has the same meaning of the turquoise line of fig. \ref{fig:fig3} but, in this case, refers to real data points, those of the Auger instrument. The ascending values of the Auger {\sc liga} energies along the green line correspond to ascending values of the ankle energies. This indicates that the energy scale of the Auger instrument is consistently moving upward, dragging on an intrinsic error in the protocol of data analysis that assigns event energies in the limited interval (0.3-5)$\times$10$^{19}$ eV.
  
The Fly's Eye, HiRes and TA data points cannot be conceived along a straight line in the E$_L$ E$_A$ plane even if they would have fortuitously laid on a straight line. In fact they are outcomes of different instruments\footnote{The HiRes I instrument derived from the ashes of Fly's Eye telescope that was dismantled after 1992, redeployed and paired at 13 km with the more powerful HiRes II in the new life of HiRes stereo born in 1999, and consequently, it is not surprising that these three instruments might have different systematic errors. Although the final exposure of HiRes stereo is modest ( $\approx$ 1500 km$^{2}$ sr year) compared to that of HiRes I ($\approx$ 4500 km$^{2}$ sr year) a variety of crucial instrumental cross-checks could be performed with HiRes I and II operated in the stereo mode. For example, using gold-plated event samples, the energy resolutions of the separate detectors HiRes I and HiRes II could be measured. The HiRes I energy resolution turned out to be $\approx$ 15 per cent according to the data of fig. 8 of \cite{soko07}.}, and hence, plausibly featured by unequal systematic errors.

In spite of that, data represented by red triangles in fig. \ref{fig:fig5} unequivocally exhibit a drift of ankle energies toward ascending values, the same tendency of the Auger data.  

\section{Driftings of the energy scales of HiRes, Auger and TA detectors 
against that of Fly's Eye}\label{sec:7drift}
The ankle energy E$_A$ in the cosmic-ray spectrum has been determined by many experiments. For example in the year 2008 the Auger Group by the words of Markus Roth \cite{roth08} affirms: ``Two spectral features are clearly visible: the so called ankle at energies of $\approx$ 3.1$\times$10$^{18}$ eV and a flux suppression above $\approx$ 3.9$\times$10$^{19}$ eV.''.

A precise measurement of the ankle energy E$_A$ has been reported in 1993 by the Fly's Eye Collaboration \cite{bird94,bird93} with a superior technique exploiting the global feature of atmospheric cascades via florescence light dodging the drawbacks of detecting charged particles on the ground. The ankle energy resulted 3.1$\times$10$^{18}$ eV \cite{bird94} with a systematic error in the energy scale of about 20 per cent. This measurement was regarded as standard for many years and this is largely documented in the literature. The ankle energy of (3.1 $\pm$ 0.5)$\times$10$^{18}$ eV measured by Fly's Eye experiment is adopted here and regarded as Premise I. In 2007 the Auger experiment found the same value for the ankle energy (see fig.10 of \cite{roth08}; also \cite{yama07}).

\begin{figure}
 \begin{center}
\includegraphics[width=0.8\textwidth]{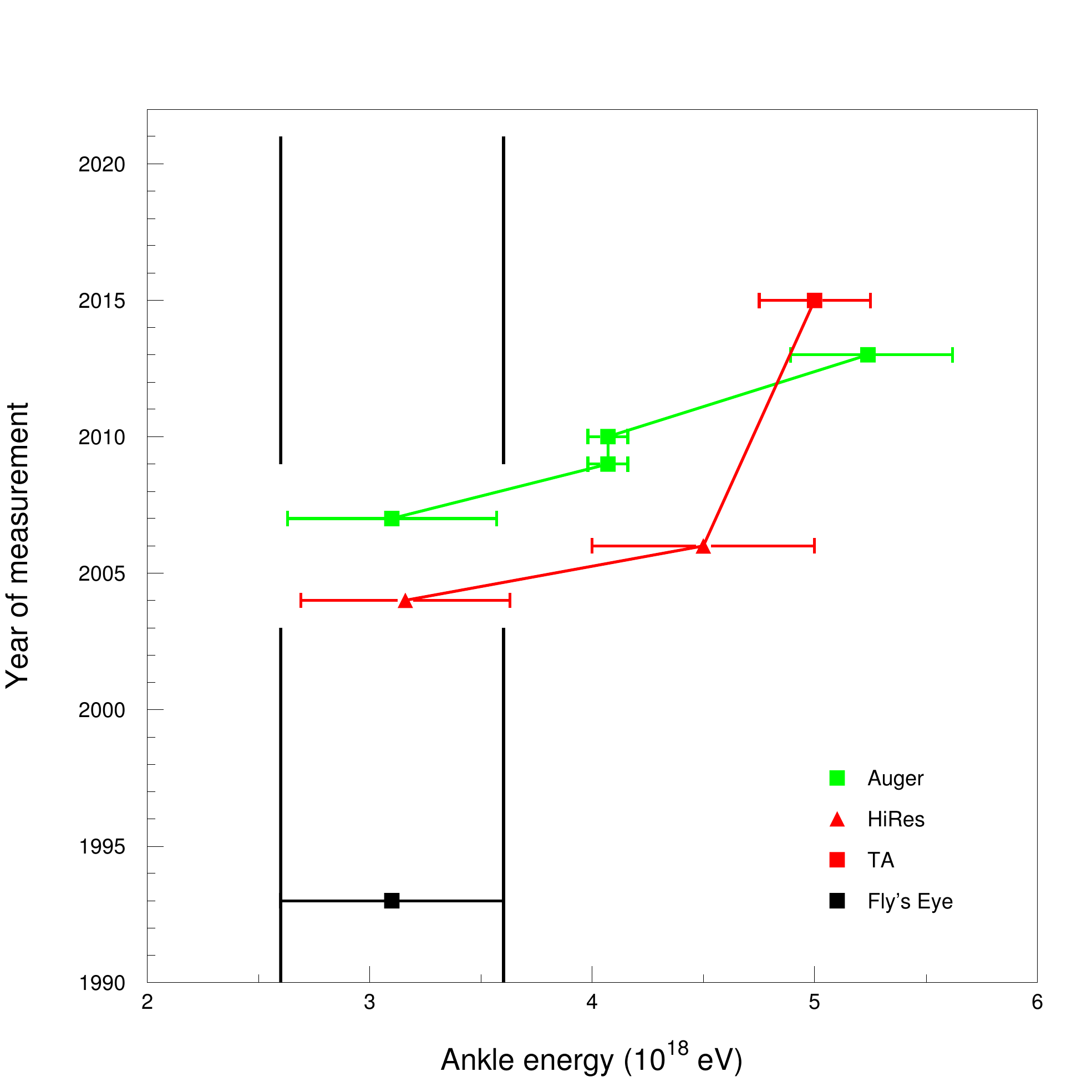}
\caption{Ankle energy versus year of the measurement for Auger (green squares), Fly's Eye (black square), HiRes (red triangles) and TA (red squares) experiments in the period 1990-2017. Patently, after 2004 ascending ankle energies are correlated with the year of the measurement regardless of the detector. On the contrary, the ankle energies remained stable for the long period 1993-2004. In the year 1993 the Fly's Eye Group made the first modern measurement of the ankle energy via florescence light released in atmospheric cascades obtaining 3.1$\times$10$^{18}$ eV \cite{bird94,bird93}. The measurement was confirmed in the subsequent year. This calorimetric method of measuring the energy of primary cosmic nuclei ultimately casted off the fetters of interpreting muon and electron patterns observed at ground. Similarly to the vertical trunk of a tree sprouting almost horizontal branches, the values of the ankle energies reported in this figure after the year 2004 expanded almost horizontally, reaching a maximum of about 5.20$\times$10$^{18}$ eV \cite{shu13}. After the year 2015 published values of the ankle energies by the Auger and TA Groups do not walk, remaining stable, close to 5.$\times$10$^{18}$ eV. \label{fig:fig5}}
\end{center}
 \end{figure}

According to the reasoning developed via the E$_L$ E$_A$ plane in the preceding Section \ref{sec:6cal}, inevitably, an artificially high value of E$_L$ energy drags on the E$_A$ energy toward an artificially high value (Premise II). From these two premises descend the following conclusions:\\ 
first, the {\sc liga} energy of (5.5 $\pm$ 0.5)$\times$10$^{19}$ eV measured by the TA Group \cite{fuku15} is too high compared to the values of (2.9 $\pm$ 0.4)$\times$10$^{19}$ eV \cite{schu09} and (4.26 $\pm$ 0.20)$\times$10$^{19}$ eV \cite{shu13} (top green square in fig. \ref{fig:fig5}) established by the Auger instrument, respectively, in the years 2009 and 2013.

Secondly, the value of the {\sc liga} energy measured in 2004 by the HiRes Group (6.3$\times$10$^{19}$ eV) \cite{thom04} is too high as well. This descends from the comparison between the ankle energy (3.1$\times$10$^{18}$ eV) established by the Fly's Eye Group \cite{bird93} and that of (4.5 $\pm$ 0.5)$\times$10$^{18}$ eV by the HiRes Group \cite{berg07}.

It is worth mentioning that the {\sc liga} energy of 2.88$\times$10$^{19}$ eV measured by the Auger experiment has been noticed and quoted by others \cite{abuz13} and it is not a biased interpretation of the energy spectrum made elsewhere \cite{cod13}.

The E$_A$ and E$_L$ energies adopted in this work (e.g. E$_A$ = 3.1$\times$10$^{18}$ eV and E$_L$ = 2.6$\times$10$^{19}$ eV) also emerge from the time evolution of the E$_A$ and E$_L$ measurements shown in figure \ref{fig:fig5}. In fact both Auger data (green squares) and the ensemble of Fly's Eye, HiRes and TA data (triangles) do exhibit common global features of ascending values of the ankle energies with the time. The time divide is the year 2004 when the evidence for the spectral break became clear and statistically solid \cite{abba04}. It is a fact exhibited by the data of fig. \ref{fig:fig5} that after the nominal date of 2004 ankle energies walk upward.

The ascending values of the ankle energy versus time of the Auger experiment indicate that only the early values \cite{yama07} are plausibly correct because the E$_A$ energy is close to 3.1$\times$10$^{18}$ eV measured by Fly's Eye Group \cite{bird94,bird93} regarded as reference value in this paper.\\

The ascending values of ankle energies (fig. \ref{fig:fig5}) concomitant with descending values E$_L$ of the {\sc liga} energy measured by Fly's Eye, HiRes and TA Groups (triangles in fig. \ref{fig:fig4}) indicate that the quantity E$_L$ - E$_A$ contracts and this behaviour is likely due to different systematic errors of the three instruments or to non linear distortions in the energy scales or to a blend of both causes. Had the three measurements (triangles) been the result of a perfect, single instrument, the physical quantity (E$_L$ - E$_A$ ) would have been 59.84, 51.3 and 49.8 in units of 10$^{18}$ eV, a thoroughly unphysical result since systematic error cancels in the difference (E$_L$ - E$_A$ ). The result would be unphysical because the quantity (E$_L$ - E$_A$ ) is constant by definition as postulated in Section \ref{sec:6cal}. But the instruments operated by the Fly's Eye, HiRes and TA Groups are indeed different.

The number of events above 10$^{20}$ eV collected by the Auger Group in the year 2007 were 3 with an estimated esposure of about 7000 km$^{2}$ sr year. Two of these three events have been rescaled in energy below 10$^{20}$ eV because they belonged to the spectrum measured by the sole surface detectors (see for example fig. 7 and 10 of \cite{roth08}).
In a comprehensive work on the energy spectrum in 2015 the number of events above 10$^{20}$ eV is 4 according to data in fig. \ref{fig:fig7} of the same paper \cite{aab15}. To-day (2017), with an exposure exceeding 42500 km$^{2}$ sr year, the number of events above 10$^{20}$ eV remains 4 according to the last minute Auger energy spectrum \cite{novo17}.\\

For comparison the number of events above 10$^{20}$ eV of the TA Group is 13 with an exposure of 8100 km$^{2}$ sr year \cite{iva17}. The collecting areas of the TA and Auger apparata are, respectively, 780 and 3000 km$^{2}$ and data taking initiated, respectively, in 2004 and 2008. In this simple arithmetic figures are the puzzling status of the data analysis of the Auger Collaboration regarding the energy spectrum around 10$^{20}$ eV
  
\section{The Yakutsk array and the {\sc liga} effect}
The outcomes of the Yakutsk Array Collaboration are difficult to examine because of the long age of the experimental facility, the complexity of the instrument and the revisions of the energy scales along the years.

The measurements of the energy spectrum of the Yakutsk Array experiment report a few events above 10$^{20}$ eV \cite{pra05}. In the very words of the YA Group in 2005 \cite{pra05}: ``\dots at present there are four events with E$_0$ $>$ 9$\times$10$^{19}$ eV that indicates to the absence of the GZK-cutoff of the spectrum but because of poor statistics and errors in energy determination this conclusion is not so reliable.''. Moreover \cite{glush14}: ``This steeping does not contradict to GZK cutoff but, probably, has a different astrophysical reason''. The steeping refers to the spectral break at 2.6 $\times$10$^{19}$ eV called in the present paper the {\sc liga} effect.

\begin{figure}
 \begin{center}
\includegraphics[width=0.8\textwidth]{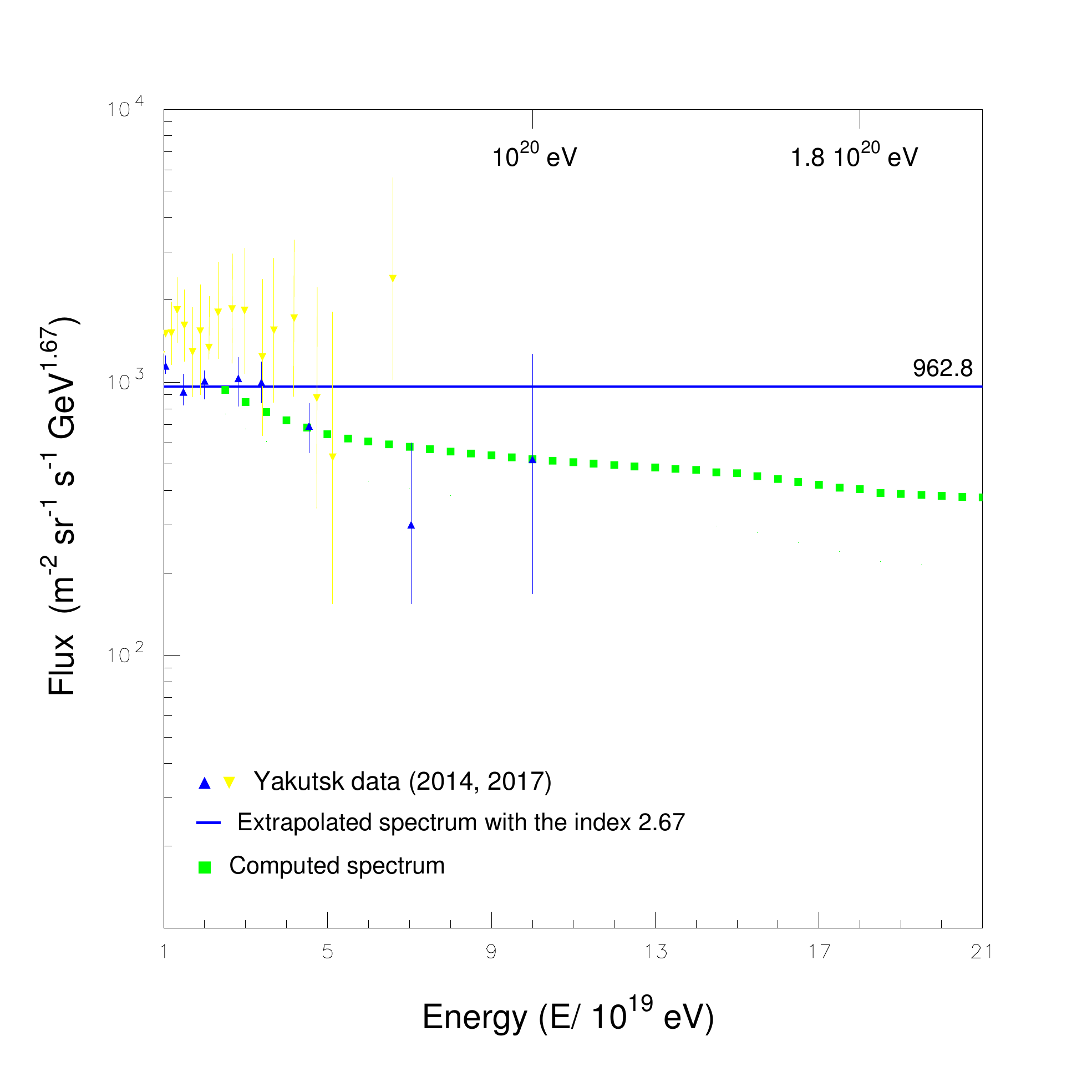}
\caption{Energy spectra measured by the Yakutsk Experiment in 2014 (blue triangles) \cite{glush14} and 2017 (yellow triangles) \cite{sabo17} compared with the predicted spectrum (tiny green squares). The horizontal blue line is the extrapolated spectrum with a constant slope of 2.67 normalized to the YA data below 3.0$\times$10$^{19}$ eV set at 962.81 particles/m$^2$ s sr GeV$^{1.67}$ corresponding to a flux of 46.00$\times$10$^{-25}$ part./m$^2$ sr s GeV at 0.721$\times$10$^{19}$ eV. This normalization is slightly different from that of figure \ref{fig:fig2} which is set at 798.17 particles/m$^2$ s sr GeV$^{1.67}$ according to the Auger data \cite{aab15}. \label{fig:fig6}}
\end{center}
 \end{figure}

The YA energy spectra \cite{glush14} along with the predicted spectrum (tiny green squares) \cite{cod17a} are shown in fig. \ref{fig:fig6}. The spectrum normalization represented by the horizontal blue line is based on YA data and amounts to 962.81 part/m$^{2}$ s sr GeV$^{1.67}$ which differs from the normalization in figure \ref{fig:fig2} of 798.17 in the same unit. The YA energy spectra change significantly in the years 2005, 2014 \cite{glush14} and 2017 \cite{sabo17}. The revisions in the energy scale of the YA instrument are the major cause for the disparity of the energy spectra in fig. \ref{fig:fig6}. The 2014 YA spectrum is quite compatible with that observed by the TA experiment, and surprisingly equal to that reported 29 years ago in 1985 (see fig4 of \cite{khris85}). Evidently, the agreement between YA data (2014, 1985) and calculation (tiny green squares) in fig. \ref{fig:fig6} is more than satisfactory but this agreement is not stringent due to the large statistical error bars and limited maximum energy.\\

The YA instrument is deployed in a valley of Lena river 55 km south of the Yakutsk city, Siberia, and started data taking in 1970 with a continuous operation up to the present times (2017). It is by far the oldest detector operating above 10$^{19}$ eV since its construction initiated in 1966. It was designed by J. B. Khristiansen, the guru of Cosmic Ray Physics who discovered with G. V. Kulikov the knee in 1958. Unlike archaic detectors the YA instrument (61.7 Nord, 129.4 East, 1020 g/cm$^{2}$, 110 meters asl) has two methods of estimating the primary particle energy: via Cherenkov light and via charged particles at ground. Notice that the YA instrument in spite of the small collecting area (see Table I) is unique and powerful because is calorimetric via Cerenkov light and detects charged particles at ground and muons underground.

In 2014 the energy scale of the YA detector has been revised \cite{glush14} and the energy spectra above 10$^{19}$ suffered downward rigid shifts in energy by a factor of 1.33 in comparison to previous reported spectra. The revision of the YA energy scale, relying upon simulations of air cascades, is anyway disquieting because, primarily, is not the result of a measurement.

The attempt made in this paper to ascertain the consistency of the energy scales of the instruments detecting cosmic rays above 10$^{19}$ eV necessarily requires an evaluation of the ankle energy (see Section \ref{sec:6cal}). In recent work of the YA Group \cite{glush14}, where a reassessment of the protocol to assign event energies is discussed, the ankle energy is not evaluated. In another work, where the YA data samples are analyzed, the ankle energy is situated around 8$\times$10$^{18}$ eV. These ankle energies are incompatible with all others determined via calorimetric measurements reported in figure \ref{fig:fig4}. 
In conclusion, since this study assumes the ankle energy of about 3.1$\times$10$^{19}$ eV with a maximum uncertainty of 25 per cent, the quoted value of $\approx$ 8$\times$10$^{18}$ eV by the Yakutsk Group \cite{ergo04} or that of (1.00 $\pm$ 0.01) $\times$10$^{19}$ eV recently estimated by a consortium \cite{daw13} are incompatible with the premise I of this work (see Section \ref{sec:7drift}).\\

Flexibility demands the vice versa: if the YA measurements of the ankle energy reported above are basically correct, the analysis of the energy scales made in this work is meaningless and misleading.

\section{The flux deficit above 10$^{20}$ \lowercase{e}V of the Auger experiment}
It is a fact that the TA and Auger flux gap in figure \ref{fig:fig1} is approximately constant in the limited band (0.3-5)$\times$10$^{19}$ eV. The difference amounts to about 20 percent (Section \ref{sec:3auta}). The revised Yakutsk energy spectrum (2014) shown in figure \ref{fig:fig6} quantitatively agrees with that of the TA experiment. Archaic detectors did measure fluxes exceeding that of the Auger Group. Calorimetric flux measurements of past detectors (Fly's Eye, Yakutsk, HiRes) also exceed the Auger flux. This global flux pattern has been already discussed in previous sections and is a prerequisite for what follows.

It is difficult to imagine that \textit{Mother Nature} distinguishes between Utah and Argentina using the directions of cosmic rays, from the galactic sources to the detectors. Also, it is unthinkable that the most sophisticated and large detector ever built might loose events at trigger level or have mundane failures in its operation. If so, the missing events above 10$^{20}$ eV in the Auger energy spectrum have to be in the collected data samples. Since aperture calculation at these extreme energies are reliable and almost constant, the protocol attributing event energies has to fail someway.

The unmotivated rescaling of the most energetic event mentioned in Section \ref{sec:5stake} indicates that event energies around 10$^{20}$ eV dance freely, up and down, in the Auger energy scale. The accordion effect, a jocose term introduced in Section \ref{sec:5stake} to designate a serious problem for the validation of the predicted spectrum \cite{cod17a}, alerts on the presence of severe distortions of the energy scales: ankles ascend the energy scale, the most energetic event descends it (see fig. \ref{fig:fig4} and \ref{fig:fig5}).

\begin{figure}
 \begin{center}
\includegraphics[width=0.8\textwidth]{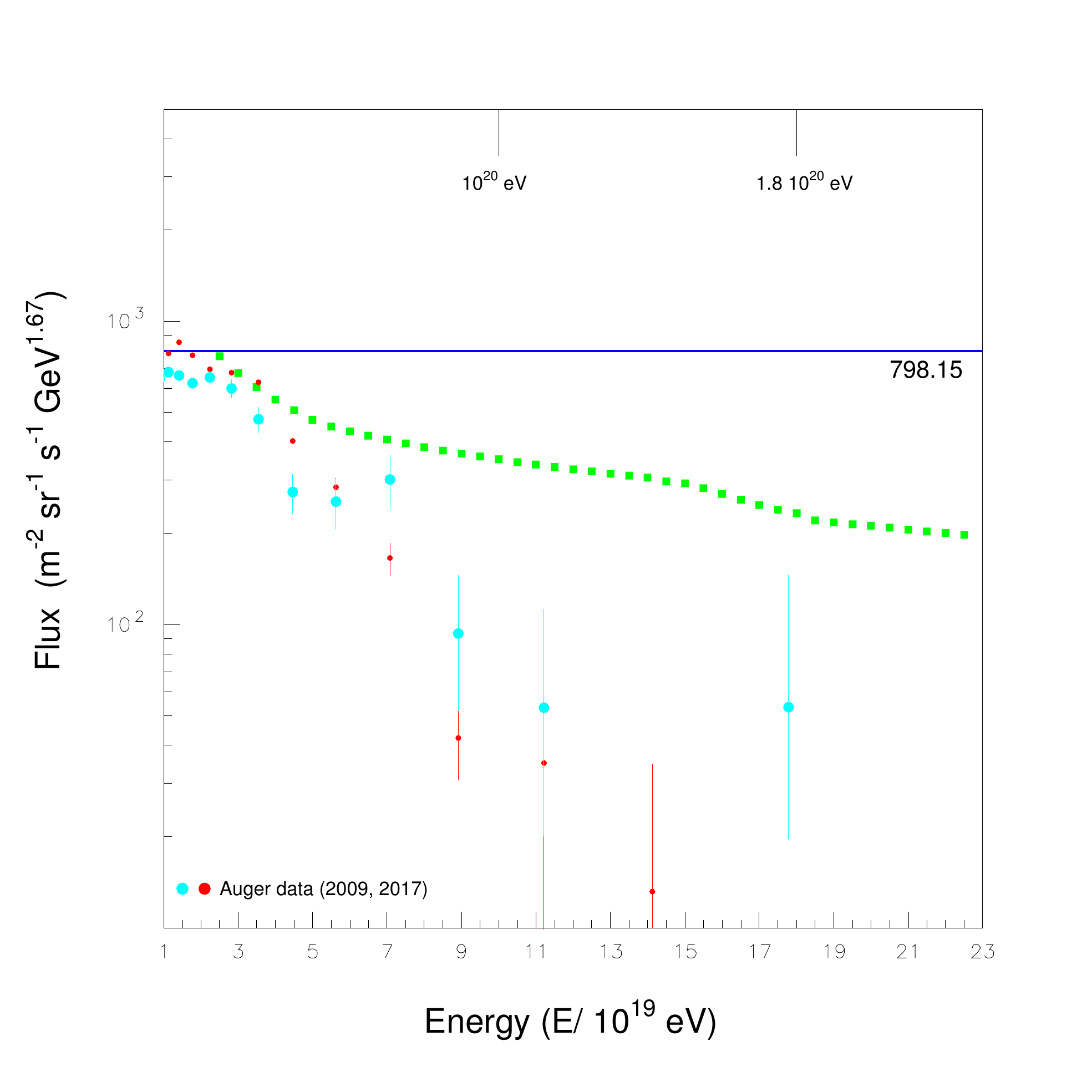}
\caption{Energy spectrum of the cosmic radiation observed by the Auger experiment in the year 2009 \cite{schu09} and that of this year 2017 \cite{novo17}. Due to the small statistical error bars there is evidence for a rotation of the energy spectrum between the year 2009 and 2017  around the energy point 5.6$\times$10$^{19}$ eV as shown in this figure and discussed in the text.\label{fig:fig7}}
\end{center}
 \end{figure}

In more practical terms: how many events from the Auger energy spectrum above 5$\times$10$^{19}$ eV have to be rescaled upward in order to obtain the acceptable flux gap of 20 percent with the TA spectrum also in the energy interval (5-25)$\times$10$^{19}$ eV?

Only a very few events, as Table 2 explicitly shows, using an arbitrary rescaling of the event energies. It is worth to recall that the flux shift of 20 per cent is within the systematic errors of TA and Auger instruments and it is a reliable reference in the interval (0.3-5)$\times$10$^{19}$ eV as already noted. The rescaling factor above the {\sc liga} energy has the form (E/E$_L$)$^{0.5}$ where E is the bin energy of the original Auger data in the range 10$^{19.35}$ - 10$^{20.15}$ eV reported this year by Vladimir Novotny \cite{novo17}. This rescaling provides 56 events above 10$^{20}$ eV, compatible with 13 events of the TA Group, being the recent TA and Auger exposures, respectively, 8100 \cite{iva17} and 42500 \cite{novo17}. After the arbitrary rescaling the spectrum relaxes and decompresses attaining a maximum energy of 4.65$\times$10$^{20}$ eV instead of 1.41$\times$10$^{20}$ eV.

\begin{table}
    \begin{center}
\caption{Number of events of the Auger energy spectrum above E$_L$=2.6$\times$10$^{19}$eV\label{tab:tab2}}
\begin{tabular}{cc|ccccccccc}

\hline
\multicolumn{2}{c|}{Bin energy}& 2.818&3.548&4.466&5.623&7.079&8.912&11.22&14.125&17.782\\
\multicolumn{2}{c|}{Rescaled energy}&2.934&4.144&5.853&8.269&11.680&16.499&23.30&32.922&46.506\\
\hline
\multirow{3}{*}{Events}&(2009)&200&110&43&28&23&5&2&0&1\\
&(2013)&676&427&188&90&45&7&3&1&0\\
&(2017)&888&569&267&130&54&10&4&1&0\\
\hline
  \end{tabular}
  \end{center}
  \end{table}

Presently, the rescaling reported in table 2 is just a product of imagination. Support to this imagination product comes from the comparison of the Auger energy spectrum of 2009 \cite{schu09} to that of 2015 \cite{aab15}, shown in figure \ref{fig:fig7}. It is evident that the 2009 energy spectrum diminished by abrupt steps as the energy increases in the interval (2.6-18)$\times$10$^{19}$ eV while that of the year 2015 decreases smoothly.

Consider  the data point in figure \ref{fig:fig7} at the energy of  5.6$\times$10$^{19}$ eV  and flux 264.4 part/m$^2$ s sr GeV$^{1.67}$. Below this energy  the 2017 flux  enhances,  above this energy decreases, and simultaneously the highest data point contracts.  The very small statistical error bars  consent to discern this extraordinary  characteristic feature of the time evolution of the Auger spectrum.  It seems that the 2017 spectrum rotates clockwise around the energy point  5.6$\times$10$^{19}$ eV with respect to the 2009 spectrum. What kind of protocol  assigning the event energies might perform such a rotation? Certainly not a rigid shift.

The sentiment prompted by the data in figure \ref{fig:fig7} is that after the year 2009 the spectrum has been hammered to appear smooth in logarithmic scales of flux and energy, plausibly within the legitimate interplay of parameters assigning event energies, but in spite of the legitimacy a potent bias operates in the penumbra. 
\begin{figure}
 \begin{center}
\includegraphics[width=0.8\textwidth]{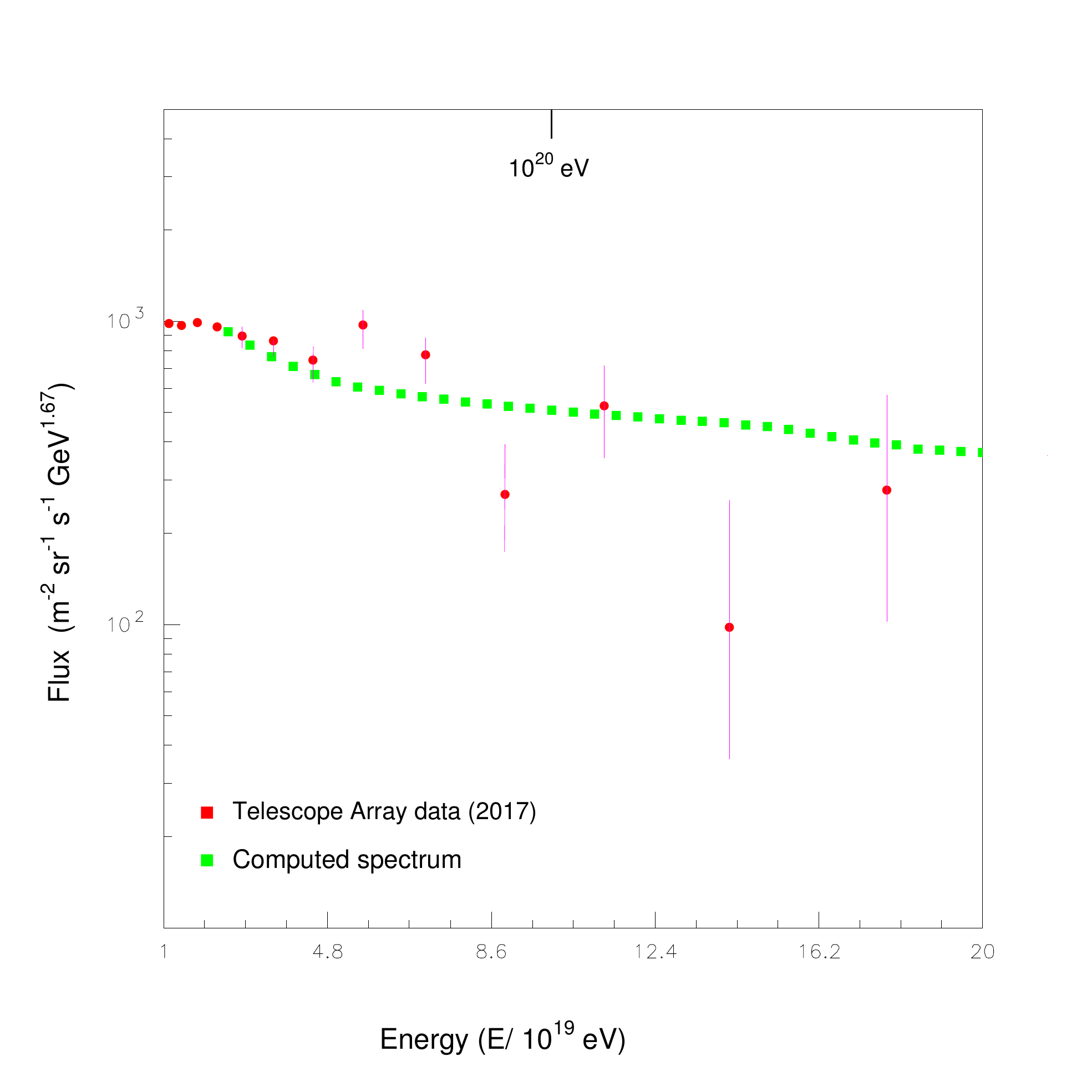}
\caption{Prediction of the energy spectrum (tiny green squares) and recent (2017) measurement of TA Collaboration around 10$^{20}$ eV.}\label{fig:fig8}
\end{center}
\end{figure}

\begin{figure}
 \begin{center}
\includegraphics[width=0.8\textwidth]{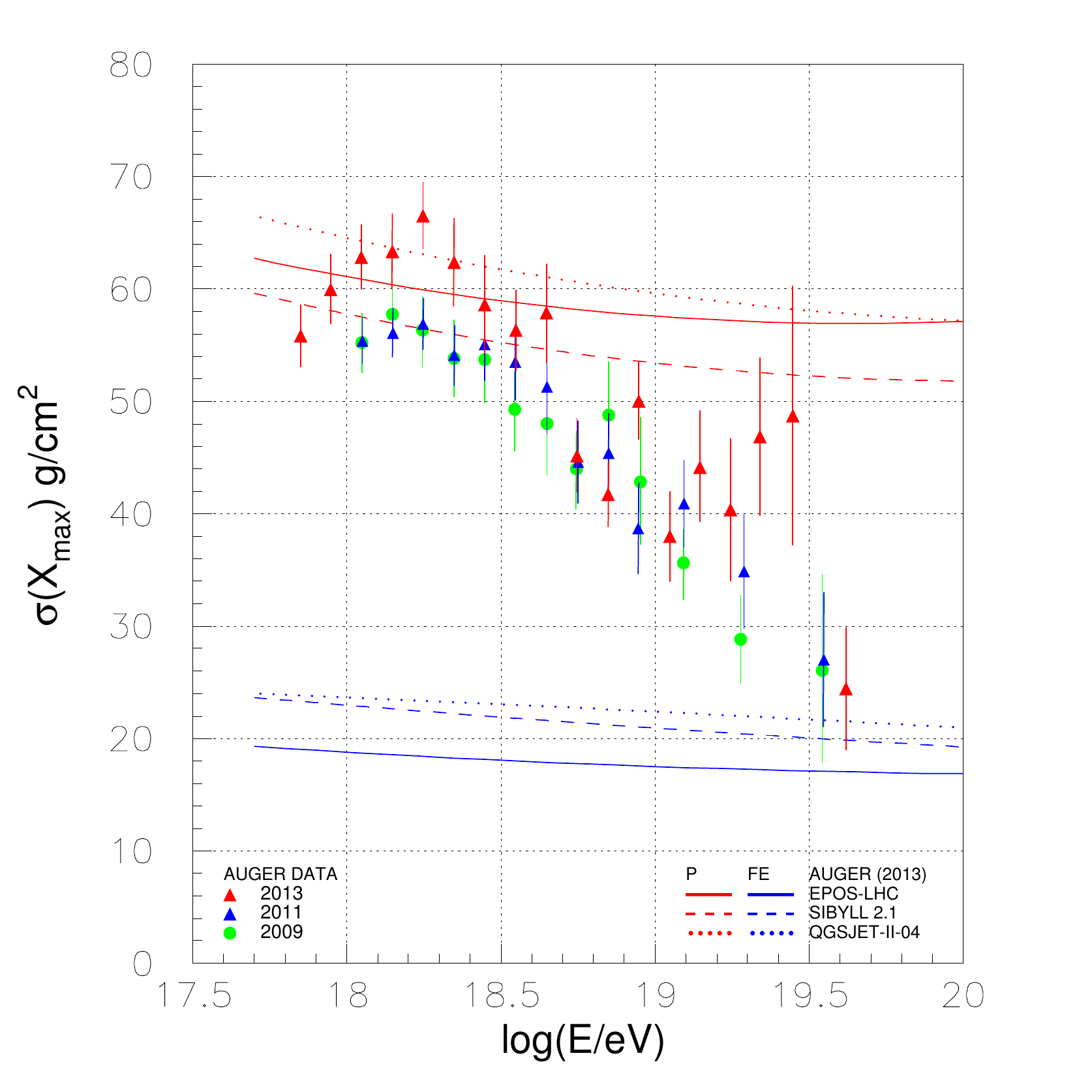}
\caption{Evidence for the step pattern in the chemical composition around 3.0$\times$10$^{19}$ eV, measured  via  the mean width of the longitudinal profile of air cascades versus energy  detected by the Auger instrument  (references to the Auger data are elsewhere \cite{cod15b}.}.\label{fig:fig9}
\end{center}
 \end{figure}

\section{Conclusions}

The prediction \cite{cod17a} of the energy spectrum of the cosmic radiation in the range 10$^{19}$ -2.4$\times$10$^{21}$ eV can be tested in the interval 10$^{19}$ -3.0$\times$10$^{20}$ eV where the flux data of 11 experiments are available (see Table I ). It results: 
\begin{itemize} 
\item[(1)] the predicted spectrum thoroughly agrees with that measured by the Telescope Array Collaboration (fig. \ref{fig:fig8} and \ref{fig:fig2}) and with trend of the chemical composition of the same experiment above 10$^{19}$ as debated in Section 3 of ref. \cite{cod17b}. 

\item[(2)] A validation of the spectrum profile in terms of the {\sc liga} effect in the range 2.6$\times$10$^{19}$ -8$\times$10$^{19}$ eV comes from the Auger data reported in the year 2009 \cite{schu09} shown in fig. \ref{fig:fig7}. The {\sc liga} effect was conceived and described in previous works \cite{cod17a,cod15a,cod15b}. The spectrum descends by steps, echoing that of fig. \ref{fig:fig3} of the prediction \cite{cod17a}, and not by a smooth profile compatible with a single slope. Presently, the precise silhouette of the spectrum is measurable only by the Auger instrument due to its overwhelming statistical precision (TA statistical errors still remain large to discern the staircase pattern). 

\item[(3)] The step pattern of the spectrum is confirmed by the step pattern of the chemical composition of the cosmic radiation.  The width of the longitudinal profile of the air cascade versus energy  measured by the Auger instrument  reported in fig. \ref{fig:fig9}, exactly around 3.0$\times$10$^{19}$ eV, exhibits an undeniable step (the H step).  At higher energy,  above 5$\times$10$^{19}$ eV, the last minute results of the Auger Group on the chemical composition  presented at ICRC 2017 by Manuela Mallamaci  (see fig. 4 and 5 of ref. \cite{malla17}) indicate  another step toward heavy nuclei\footnote{According to the Author of this paper the outstanding results of  $ln(A)$ versus energy reported in fig. \ref{fig:fig4} of the quoted paper \cite{malla17} have been belittled and masqueraded as Monte Carlo exercises by the Auger Collaboration.}.
  
\item[(4)] There is good agreement between computed and observed spectra with the ensemble of the measurements except with those of the Auger Group around 10$^{20}$ eV. Good agreement here means that spectrum data of the eleven experiments are erratically scattered above and below the prediction in the range 2.6$\times$10$^{19}$ -3$\times$10$^{20}$ eV. The Auger absolute flux is severely deficient above 10$^{20}$ eV. It disagrees with the predicted spectrum \cite{cod17a} and with the fluxes of all other calorimetric experiments (TA, HiRes, Fly's Eye, Yakutsk). Notice that the Auger spectrum profile mentioned in (2) refers to the interval (1-8)$\times$10$^{19}$ eV while the flux deficiency becomes intolerable in the interval (8-20)$\times$10$^{19}$ eV. 
\end{itemize} 

By assuming that the energy scales of different instruments is the dominant cause of the flux gap between Auger spectrum and those of the other calorimetric experiments, several major inconsistencies regarding the Auger energy scale emerge:
\begin{itemize}
\item[(A)] ascending values of the ankle energies, E$_A$, with the year of the measurement (see fig. \ref{fig:fig5}).

\item[(B)] Ascending values of the {\sc liga} energies, E$_L$, correlated to ascending ankle energies 
   (fig. \ref{fig:fig4}); this correlation provides evidence for the drifting of the energy scale of the 
   Auger experiment along the years.

\item[(C)] The unmotivated rescaling of the most energetic event from the energy 1.77$\times$10$^{20}$ eV
  down to 1.41$\times$10$^{20}$ eV. This rescaling has fundamental implications in Cosmic Ray Physics.
  In fact the energy of 1.77$\times$10$^{20}$ eV and the staircase profile of fig. \ref{fig:fig7} obliterate the
   relevance of any statistical tests with a single slope on the existence of the GZK effect.
\end{itemize}

Notice further that the fictitious GZK effect is comfortably discarded by the heavy chemical composition above 10$^{19}$ eV reported in 2007 by the Auger Collaboration and confirmed by the TA Group in recent years (see Section 3 of ref. \cite{cod17b}; also \cite{cod13}).


\begin{thebibliography}{23}
\addcontentsline{toc}{section}{References}
{
\expandafter\ifx\csname natexlab\endcsname\relax\def\natexlab#1{#1}\fi
\providecommand{\bibinfo}[2]{#2}
\ifx\xfnm\relax \def\xfnm[#1]{\unskip,\space#1}\fi
\bibitem{cunn80} Cunningham G., et al. (Haverah Park Coll.) 1980, \textit{The energy spectrum and  arrival direction distribution of cosmic rays with energies above 10$^{19}$ electronvolts}, The Astrophysical Journal, 236, L71-L75.
\bibitem{winn86} Winn M.~M., et al. (SUGAR Coll.) 1986, \textit{The cosmic ray spectrum above 10$^{17}$ eV}, \ Journal of Physics G, 12, 653-674.
\bibitem{tak02} Takeda M., et al. (Akeno-AGASA Coll.) 2002, \textit{Energy determination in the Akeno Giant Air Shower Array  Experiment}, astro-ph/0209422v3, 13 November 2002; also Astroparticle Physics 2003.
\bibitem{shino06} Shinozaki K. (AGASA Coll.) 2006, \textit{Observation of UHECRs by AGASA} (oral presentation), Quarks-~2006, St. Petersburg, 19-25 May 2006.
\bibitem{abba04} Abbasi R.~U. et al. (HiRes Coll.) 2004, \textit{First Observation of the Greisen-Zatsepin-Kuzmin Suppression}, Physical Review Letters, 92, 15, 2004; also astro-ph/0703099v2.
\bibitem{yama07} Yamamoto T. (Auger Coll.) 2007, \textit{The UHECR spectrum  measured at the Pierre Auger Observatory and its astrophysical implications}, Proceedings 30th ICRC 2007; also arXiv: 0707.2638v1.  
\bibitem{mat11} Matthews N.~L. (TA Coll.) 2011, \textit{The Telescope Array Experiment}, 32th ICRC paper 1306.
\bibitem{law91} Lawrence M.~A., Reid R.~J.~O. and Watson A.~A. (1991), \textit{The cosmic ray energy spectrum above 4$\times$10$^{17}$ eV as measured by the Haverah Park array}, Journal of Physics G, 17, 733-757; also 21th ICRC 1990, 3, 159-163.
\bibitem{schu09} Schuessler M.~V. (Auger Coll.) 2009, \textit{Measurements of the cosmic ray energy spectrum  above 10$^{18}$ eV with the Pierre Auger Observatory}, 31th ICRC  \L\'od\'z, Poland.
\bibitem{jui16} Jui C.~H. (TA Coll.) 2016, \textit{Results from the Telescope Array experiment}, Proceedings of Nuclear and Particle Physics, 273, 445-448.
\bibitem{cod13} Codino A. (2013), \textit{The absence of the GZK depression in the energy  spectrum of the cosmic radiation}, 33th ICRC, Rio de Janeiro, Brasil.
\bibitem{cod17a} Codino A. (2017), \textit{About the energy interval above the ankle where the cosmic radiation consists only of ultraheavy nuclei from the Zinc to the Actinides}, Journal of Applied Mathematics and Physics, 5, 225-237.
\bibitem{cod15a} Codino A. (2015), Slides at the CRIS Conference,\\ \url{https://agenda.infn.it/getFile.py/access?contribId=7&sessionId=4&resId=0&materialId=slides&confId=8818}, 14-16 September, Gallipoli, Italy.
\bibitem{cod17b} Codino A. (2017), \textit{The energy spectrum of ultraheavy nuclei above 10$^{20}$ eV}, Journal of Applied Mathematics and Physics, 5, 8, 1540; also arXiv: 1707.02487v1 astro-ph.HE, 8 July 2017.
\bibitem{sala11}  Salamida F. (Auger Coll.) 2011, \textit{Update of the measurement of the CR energy spectrum above 10$^{18}$ eV made using the Pierre Auger Observatory}, 32th ICRC Beijing, China.
\bibitem{aab15} Aab A. (Auger Coll.) 2015, \textit{Measurements of the cosmic ray spectrum above 4$\times$10$^{18}$ eV using inclined events detected  with the Pierre Auger Observatory}, arXiv: 1503.07786v1 astro-ph-HE.
\bibitem{tin14} Tinyakov P. et al. (TA Coll.) 2014, \textit{Latest Results from the Telescope Array}, Nuclear Instruments and Methods A, 742, 29-34.
\bibitem{sabo17} Sabourov A. from the Yakutsk Collaboration, private communication, 2017.
\bibitem{cod15b} Codino A. (2015), \textit{The Knee and Ankle Feature derived from the principle of constant Indices and the Galactic Accelerator}, 34th ICRC, paper 466, La Hague, Holland.  
\bibitem{fuku15} Fukushima M. (TA Coll.) 2015, \textit{Recent results from Telescope Array}, arXiv: 1306.6138v1 astro-ph.HE, 24 March 2015.
\bibitem{bird94} Bird D.~J. et al. (Fly's Eye Coll.) 1994, \textit{The cosmic-ray energy spectrum observed by Fly's Eye}, The Astrophysical Journal, 424, 491-502.   
\bibitem{bird93} Bird D.~J. et al. (HiRes Coll.) 1993, \textit{Evidence for Correlated Changes in the Spectrum and Composition of Cosmic Rays at extremely High Energy}, Physical Review Letters, 71, 3401-3404.
\bibitem{cod15c} Codino A., Progress and Prejudice in Cosmic Ray Physics until  2006, (Societ\`a Editrice  Esculapio, Bologna, Italy), 2015 (\url{http:/www.editrice-esculapio.com/codino-progress-and-prejudice-in-cosmic-ray-physics-until-2006/}).
\bibitem{maris14} Maris I.~C. et al. (Auger and TA Consortium) 2014, \textit{The energy spectrum of ultra high energy cosmic rays}, Proceedings of the Int. Symp. of UHERC2014, Springdale, USA.
\bibitem{daw13} Dawson B.~R. et  al., 2013, \textit{The energy spectrum of cosmic rays at the highest energies}, arXiv: 1503.0696v1 astro-ph.. HE, 26 June 2013.  
\bibitem{maria11} Mariazzi A. (Auger Coll.) 2011, \textit{A new method for determining the primary energy from the calorimetric showers observed in hybrid mode on a shower-by-shower basis}, 32th  ICRC, Beijing, China.
\bibitem{verzi} Verzi V. (Auger Coll.) 2013, \textit{The Energy Scale of the Pierre Auger Observatory}, 33th ICRC, paper 928, Rio de Janeiro, Brasil.
\bibitem{tak03} Takeda M. (AGASA Coll.) 2003, \textit{Energy determination in the Akeno Giant Air Shower Array experiment}, Astroparticle Physics, 19,  447-462.
\bibitem{abba05} Abbasi R. U. et al. (HiRes Coll.) 2005,  \textit{Observation of the ankle and Evidence for a High-Energy break in the cosmic ray spectrum},  arXiv: astro-ph/0501317v2, 5 June 2005 (14 pages).
\bibitem{glush14} Glushkov A.~V. et al. (YA Coll.) 2014, \textit{Revision of the energy calibration of the Yakutsk EAS array}, arXiv: 1408.6302v1 astro-ph-HE, 27 August, 2014.
\bibitem{lin61} Linsley J., Scarsi L., Rossi B. (Volcano Ranch Coll.) 1961,  \textit{Extremely energetic cosmic-ray event}, Physical Review Letters,  6 pages, 485-487.
\bibitem{bow83} Bower A.~J. et al., \textit{On estimating the energy of giant air-shower primaries}, Journal of Physics  G,  9, L53-L58, 1983.
\bibitem{ave02} Ave M. et al. (Haverah Park Coll.) 2002, \textit{Reassessment of the Haverah Park Energy Spectrum above 3$\times$10$^{17}$ eV}, astro-ph/0112253v2, 11 March 2002; also Astroparticle Physics 2003.
\bibitem{khris85} Khristiansen G.~B. (1985), \textit{} Proceedings of 19th ICRC, 9, 487.
\bibitem{shu13} Schulz A. (Auger Coll.) 2013, \textit{The measurements of the energy spectrum of the cosmic rays above 3$\times$10$^{17}$ eV with the Pierre Auger Observatory}, Proceedings 33th ICRC, Rio de Janeiro, Brasil.
\bibitem{barci11} Barcikovsky E.~L., Thesis work (2011), \textit{The composition of ultra high energy cosmic rays through hybrid analysis at Telescope Array}, Department of Physics and Astronomy, University of Utah, Utah, December 2011.
\bibitem{soko07} Sokolsky P. (HiRes Coll.) 2007, \textit{Recent results from ther High Resolution Fly's Eye experiment: an introduction}, Proceedings of CRIS 2006, Nuclear Physics B, 11.
\bibitem{roth08} Roth M. (Auger Coll.) 2007, \textit{Measurement of the energy spectrum from the Pierre Auger Experiment}, Proceedings of CRIS 2007, Nuclear Physics B, 12-18.
\bibitem{thom04} Thomson G. (HiRes Coll.) 2004, \textit{New results from HiRes experiment}, Nuclear Physics B, 136, 28-33.
\bibitem{berg07} Bergman D.~R. (HiRes Coll.) 2007, \textit{Observation of the GZK cutoff using the HiRes detector}, Proceedings of CRIS 2006, Nuclear Physics B, 19-26.
\bibitem{abuz13} Abu-Zayyad, T. et al. (2013),\textit{} The Astrophysical Journal Letters, 768, L1.
\bibitem{novo17} Novotny V. (Auger Coll.) 2017, \textit{Energy spectrum  of cosmic rays measured with the Pierre Auger Observatory}, 52th Rencontres de Moriond, La Thuile, 25-28 March 2017.
\bibitem{iva17} Ivanov D. (TA Coll.), 2017, private communication.  
\bibitem{pra05}  Pravdin M.~V. et al. (YA Coll.) 2005, \textit{Estimation of giant shower energy at the Yakutsk  EAS array}, 29th  ICRC, 7,  pages 243-246.
\bibitem{ergo04} Egorova V.~P. et al. (YA Coll.) 2004, \textit{The spectrum features of UHECR below and surrounding the GZK}, Proceedings of CRIS 2004, Nuclear Physics B, 136, 3-11. 
\bibitem{malla17} Mallamaci M. (Auger Coll.) 2017, \textit{Measurements of the depth of maximum muon production and of its fluctuations in extensive air showers above 1.5$\times$10$^{19}$eV at the Pierre Auger Observatory}, Proceedings of 35th ICRC, paper 509, Busan.
}
\end{thebibliography}
\end{document}